# Tailoring Dzyaloshinskii-Moriya interaction in a transition metal dichalcogenide by dual-intercalation

Guolin Zheng[1#], Maoyuan Wang[2,3,4#], Xiangde Zhu[5#], Cheng Tan[1], Jie Wang[5,6],

Sultan Albarakati[1], Nuriyah Aloufi[1], Meri Algarni[1], Lawrence Farrar[1], Min Wu[5],

Yugui Yao[2,3], Mingliang Tian[7,5,8*], Jianhui Zhou[5*], Lan Wang[1*]

[1]School of Science, RMIT University, Melbourne, VIC 3001, Australia.

[2]Centre for Quantum Physics, Key Laboratory of Advanced Optoelectronic Quantum Architecture and Measurement (MOE), School of Physics, Beijing Institute of Technology, Beijing, 100081, China.

[3]Beijing Key Lab of Nanophotonics & Ultrafine Optoelectronic Systems, School of Physics, Beijing Institute of Technology, Beijing, 100081, China.

[4]International Center for Quantum Materials, School of Physics, Peking University, Beijing 100871, China.

[5]Anhui Province Key Laboratory of Condensed Matter Physics at Extreme Conditions, High Magnetic Field Laboratory, HFIPS, Chinese Academy of Sciences (CAS), Hefei 230031, Anhui, China.

[6]University of Science and Technology of China, Hefei 230026, Anhui, People's Republic of China.

[7]Department of Physics, School of Physics and Materials Science, Anhui University, Hefei 230601, Anhui, China.

[8]Collaborative Innovation Center of Advanced Microstructures, Nanjing University, Nanjing 210093, China.

[#] Those authors equally contribute to the paper.

[*] Corresponding authors. Correspondence and requests for materials should be addressed to J. Z. (email: jhzhou@hmfl.ac.cn); M. T. (email: tianml@hmfl.ac.cn); L. W. (email: lan.wang@rmit.edu.au).

# Abstract

Dzyaloshinskii-Moriya interaction (DMI) is vital to form various chiral spin textures, novel behaviors of magnons and permits their potential applications in energy-efficient spintronic devices. Here, we realize a sizable bulk DMI in a transition metal dichalcogenide (TMD) 2H-TaS$_2$ by intercalating Fe atoms, which form the chiral supercells with broken spatial inversion symmetry and also act as the source of magnetic orderings. Using a newly developed protonic gate technology, gate-controlled protons intercalation could further change the carrier density and intensely tune DMI via the Ruderman-Kittel-Kasuya-Yosida mechanism. The resultant giant topological Hall resistivity $\rho_{xy}^T$ of $1.41\ \mu\Omega \cdot cm$ at $V_g = -5.2\ V$ (about 424% larger than the zero-bias value) is larger than most known chiral magnets. Theoretical analysis indicates that such a large topological Hall effect originates from the two-dimensional Bloch-type chiral spin textures stabilized by DMI, while the large anomalous Hall effect comes from the gapped Dirac nodal lines by spin-orbit interaction. Dual-intercalation in 2H-TaS$_2$ provides a model system to reveal the nature of DMI in the large family of TMDs and a promising way of gate tuning of DMI, which further enables an electrical control of the chiral spin textures and related electromagnetic phenomena.

**Introduction.**

The marriage of the broken local spatial inversion symmetry (SIS) and strong spin-orbit coupling (SOC) in magnetic materials could lead to the asymmetric exchange interaction, Dzyaloshinskii-Moriya interaction (DMI) [1, 2]. DMI has attracted increasing attention due to its capability to stabilize the chiral spin textures, such as magnetic skyrmions, chiral domain walls [3-5], and realize the novel physics of elementary excitations in magnetic insulators, including the spin Nernst Hall effect [6], the thermal Hall effect of magnons [7, 8]. Passing through the chiral spin textures, electrons can feel an artificial gauge field and accumulate a finite real-space Berry phase [9], resulting in a remarkable transport signature-topological Hall effect (THE) [10, 11], which acts the probe of chiral spin textures and the underlying DMI. Thus controlling DMI would greatly facilitate the manipulation of chiral spin textures and the related anomalous electromagnetic responses as well as their potential application in energy-efficient spintronic devices. DMI in bulk materials usually originates from the inversion asymmetry in the natural unit cells of crystals [12-14], or the structural inhomogeneity along the thickness direction, as in amorphous ferrimagnets GdFeCo [15]. In contrast, realizing DMI in layered materials permits the manipulation of chiral spin textures and investigation of various related fascinating physics in few atomic layers. While in a large family of layered materials, SIS is respected in the natural unit cell and usually forbids DMI, we, however, demonstrate a promising way to induce sizable DMI by breaking SIS in the enlarged supercell through alternatively intercalating heavy magnetic atoms also as the source of magnetic orderings.

The intercalation of magnetic atoms (V, Cr, Mn, Fe, Co, Ni) in TMDs 2H-TaS$_2$ and -NbS$_2$ leads to various magnetic ground states (including easy axis/plane ferromagnetism (FM) and antiferromagnetism (AFM)) [16, 17] and novel strongly correlated states [18], providing a leading-edge field searching for DMI and potential chiral magnetic structures. It has been reported that magnetic fields could turn the magnetically intercalated TMDs into novel chiral spin textures, such as the chiral solitons and chiral conical states in Cr$_{1/3}$NbS$_2$ [19, 20], the chiral domain walls in Fe$_{1/3}$TaS$_2$ [21]. Although DMI is crucial to these fascinating experiments above, the existence and nature of DMI in magnetically intercalated TMDs were, however, unknown. In addition, DMI in itinerate intercalated TMDs is usually dominated by the Ruderman-Kittel-Kasuya-Yosida (RKKY) interaction as the FM interactions [22] and can in principle be controlled by tuning the carrier density. However, an electrical tuning of DMI so as to all electrically control the chiral spin textures in itinerate magnets is still a big challenge. This is because for conventional gate technology, electric-field tends to be greatly screened in itinerant magnets by mobile carriers, which cannot effectively modulate the carrier density and ferromagnetism. Ionic liquid (Li$^+$) [23] can accumulate substantial charge carriers. However, it can only tune the carriers close to the surface [24].

In this article, we demonstrate that DMI can be induced and controlled in a TMD 2H-TaS$_2$ by dual-intercalation. Intercalating Fe atoms into 2H-TaS$_2$, Fe$_{1/3-\delta}$TaS$_2$ ($\delta \leq$ 0.05), DMI is confirmed by the observation of THE at low temperatures. Moreover, protons intercalation induced by electrical gating could further change the carrier density then largely tune DMI via the RKKY mechanism, resulting in a huge topological Hall resistivity of $1.4\,\mu\Omega\cdot cm$ at $V_g = -5.2\,V$. Theoretical analysis shows that this large THE is attributed to the 2D Bloch-type spin textures (skyrmions or chiral domain walls) stabilized by the large DMI that comes from the presence of the chiral supercells and strong SOC. Direct evaluation of the anomalous Hall conductivity (AHC) and Berry curvature reveals the origin of the large AHC in experiments. Tailoring DMI in 2H-TaS$_2$ by dual-intercalation may reveal the universality of DMI and open up the opportunity of more investigations of chiral spin textures in a large family of TMDs.

**Results.**

**Characterization of the magnetic properties**. Intercalation of Fe atoms in 2H-TaS$_2$ possesses a large perpendicular magnetic anisotropy (PMA) [16, 25]. Besides, in Fe$_x$TaS$_2$ with moderate Fe concentrations ($0.28 \leq x \leq 0.33$), Fe atoms intercalate between layers and can form $\sqrt{3}a \times \sqrt{3}a$-type supercell ($a$ is the hexagonal lattice parameter of 2H-TaS$_2$) [26, 27], as shown in Fig. 1a and b. This resulting chiral $\sqrt{3}a \times \sqrt{3}a$-type supercell harbouring strong SOC of Ta and Fe atoms and in the absence of the SIS [21], allows for a sizable DMI [3-5]. To verify this point explicitly, we first focus on the transport properties in Fe$_{0.28}$TaS$_2$. Fig. 1c shows the temperature dependent Hall resistivity in sample S1 with thickness of $80\,nm$. In conventional FM metals, the Hall resistivity $\rho_{xy}$ has two components, the normal Hall resistivity $\rho_{xy}^N$ due to Lorentz force induced by an external magnetic field and the anomalous Hall resistivity $\rho_{xy}^A$ scaling with magnetization [28]. This picture is in line with our observations above $20\,K$, where Hall resistivity exhibits a nearly square-shaped hysteresis loop with a sharp transition near the coercive field $\mathbf{B_c}$ (Figure S2 in SI). While below $20\,K$, an extra "hump" in $\rho_{xy}$ near the coercive field emerges, which is not proportional to the magnetization process and is usually attributed to the THE ($\rho_{xy}^T$) induced by unconventional spin textures, as shadowed by the light purple in Fig. 1b. The total Hall resistivity now consists of three parts: $\rho_{xy} = \rho_{xy}^N + \rho_{xy}^A + \rho_{xy}^T$. In order to obtain the topological Hall resistivity component, we linearly fitted the Hall resistivity at high field to subtract the normal Hall component $\rho_{xy}^N$. Due to the large PMA, the anomalous Hall resistivity of $Fe_{0.28}TaS_2$ exhibits square-shaped hysteresis loops. Thus the anomalous Hall part can be subtracted by linearly extrapolating the high field anomalous Hall resistivity over the coercive field $\mathbf{B_c}$, the obtained hump structures are ascribed to topological Hall resistivity.

Fig. 1d shows the extracted $\rho_{xy}^T$ of sample S1 at various temperatures. $\rho_{xy}^T$ decreases with increasing temperatures, and it drops to zero while above $20\,K$. As discussed above, $\sqrt{3}a \times \sqrt{3}a$-type supercell widely exists in $Fe_xTaS_2$ with different Fe doping concentrations ($0.28 \leq x \leq 0.33$). Hence DMI can be developed in $Fe_xTaS_2$ in this range of Fe doping level. Beside $Fe_{0.28}TaS_2$, we carried out extra electric transport measurements in crystals with higher Fe concentrations ($Fe_{0.3}TaS_2$). As expected, a large THE was also observed in $Fe_{0.3}TaS_2$ (Figure S3 in SI). The observation of THE in $Fe_xTaS_2$ is a direct transport evidence for DMI in Fe intercalated $2H$-$TaS_2$.

Another electromagnetic response induced by intercalation is large anomalous Hall effect. The spontaneous FM order of intercalated Fe atoms with strong SOC also acts the source of AHE [9]. Fig. 2a presents temperature dependent AHC, $\sigma_{xy}^A = \rho_{xy}^A/\left({\rho_{xy}^A}^2 + \rho_{xx}^2\right)$ in sample S1. At $T = 2\,K$, $\sigma_{xy}^A$ reaches $478\,\Omega^{-1}cm^{-1}$, and it drops to $275\,\Omega^{-1}cm^{-1}$ at $T = 50\,K$. By using the anomalous Hall angle $\theta = \sigma_{xy}^A/\sigma_{xx}$ ($\sigma_{xx}$ is the longitudinal conductivity) to measure the contribution of anomalous Hall current with respect to the normal current [9], we find that the anomalous Hall angle in S1 is as large as 5% (Figure S4 in SI). To fully understand the intrinsic AHC induced by the Berry curvatures of electrons in Fe$_{1/3-\delta}$TaS$_2$ ($\delta \leq 0.05$), we get the band structure of Fe$_{1/3}$TaS$_2$ through the First-principles calculations. Due to the spontaneous magnetization of Fe atoms, the spin degeneracy of the band structure is broken, splitting the bands of different spins. The different effective masses of hole pockets with different spins cross each other around the Fermi energy, forming nodal lines in quantity at different k$_z$ planes of the Brillouin zone. The space group P6$_3$22 of the Fe$_{1/3}$TaS$_2$ allows nonsymmorphic protected spin-polarized Weyl points in the $\boldsymbol{\Gamma - A}$ direction. Since these Weyl points are far away from the Fermi level, they would not dominate the intrinsic AHC. When the SOC is taken into account, most of the nodal lines will open gaps, contributing AHC through the Berry curvature. Fig. 2c displays the distribution of the Berry curvature on the $\mathbf{k_x - k_y}$ plane with $\mathbf{k_z} = 0$ at $E_F = 0$. Although the gapped nodal lines are completely not in the same energy, the distribution of the Berry curvature forms several circles, which results in a large intrinsic AHC $\sigma_{xy,in}^A$, as shown in Fig. 2d. The calculated intrinsic AHC $\sigma^A_{xy,in}$ is about $400\ \Omega^{-1}cm^{-1}$ at $E_F = 0$, quantitatively comparable with the experimental results. We also consider the scaling relation between AHC and $\sigma_{xx}$ ($\sigma^A_{xy} \propto (\sigma_{xx})^\alpha$). Plotting $\sigma^A_{xy}$ vs $\sigma_{xx}$ against temperature in Fig. 2e, we find the scaling exponent $\alpha \approx 1.4$, close to the value of 1.6 for the intrinsic AHC in multiband disordered metals [9]. It is consistent with the complex multiband structure as shown in Fig. 2b.

**Protonic gating.** Now we focus on controlling of DMI by gate-induced protons intercalation. Compared with widely used Lithium ions, protons are more movable and controllable by gating due to much smaller size, allowing for a large modulation of charge carriers and the magnetic interactions (such as FM and DMI) in bulk itinerate magnets. To achieve this, we developed a new protonic gate (Fig. 3a, see methods), and find that both the observed THE and anomalous Hall effect (AHE) can be dramatically modulated by gate-controlled protons intercalation, suggesting high tunability of DMI. Fig. 3b exhibits the gate-tuned topological Hall and anomalous Hall resistivity $\rho^T_{xy} + \rho^A_{xy}$ in sample S2 at $8\ K$ with a thickness of $115\ nm$. At $V_g = 0\ V$, when the magnetic field is swept between $-7\ T$ and $+7\ T$, a large THE (as shadowed by the light purple colour) appears around $\pm\ 3\ T$. Increasing the voltage from $0\ V$ to $-5.2\ V$, we find that both anomalous Hall resistivity and topological Hall resistivity enhanced with increasing gate voltages. Note that, the coercivities keep unchanged during the whole gating process, that is the PMA is almost unchanged in this process, despite both THE and AHE can be dramatically tunned. On the other hand, the stabilization of chiral spin textures is determined by the competition between PMA and DMI, thus the unvaried coercivities indicate that the gate-tuned THE is mainly ascribed to the change of DMI under various gate voltages. Fig. 3c shows the gate-dependent amplitudes of the topological and anomalous Hall resistivity. As we can see, the topological Hall resistivity changes from $0.269\ \mu\Omega \cdot cm$ at $0\ V$ to $1.41\ \mu\Omega \cdot cm$ at $-5.2\ V$. Note that such a huge topological Hall resistivity ($\rho_{xy}^T$) at $V_g = -5.2\ V$ is larger than most of the known magnetic systems, which is almost the largest one observed in chiral magnets so far (table 1 in SI). Simultaneously, the anomalous Hall resistivity ($\rho_{xy}^A$) monotonically changes from $1.38\ \mu\Omega \cdot cm$ ($0\ V$) to $4.6\ \mu\Omega \cdot cm$ ($-5.2\ V$). The variation of the topological Hall resistivity ($\Delta\rho_{xy}^T = \rho_{xy}^T(-5.2V) - \rho_{xy}^T(0\ V)$) normalized by the zero-bias value is as large as 424% (233% for $\rho_{xy}^A$), which is much larger than the one (~55%) in oxide heterostructures of $SrRuO_3$-$SrIrO_3$ tuned by the applied electric field [29]. Note that both anomalous Hall and topological Hall exhibit similar gate-dependence, this is due to their similar dependence on gate-induced carrier density. The evolution of gate-dependent anomalous Hall effect in Fig. 3c (also in Figure S9 and Figure S10 in SI) can be fairly well captured by our simulation of AHC in Fig. 2d. The intense modulation of THE qualitatively demonstrates a large and promising electrical tailoring of DMI, which has not been achieved so far. Gate-induced protons intercalation thus may provide a powerful way of electric control of transport phenomena in spintronic and electronic devices with large charge densities.

**Discussion**

To further confirm the DMI in intercalation-induced chiral supercells, we consider the impacts of crystal symmetry breaking on the spin textures. First, the large PMA implies an anisotropic ferromagnet, leading to merely two-fold degenerate ground states (Ising-type state). Second, $Fe_{1/3}TaS_2$ has a layered hexagonal structure of 2H-type $TaS_2$ intercalated by Fe atoms which belongs to the non-centrosymmetric chiral space group $P6_322$ [21]. That is, the Fe atoms intercalating in $\sqrt{3}a \times \sqrt{3}a$-type supercell will break the SIS. As a result, a substantial DMI is allowed due to the combination of strong SOC of Fe and Ta atoms and the broken SIS [3-5]. This DMI may share the same RKKY mechanism as the out-of-plane FM ordering demonstrated in $Fe_{1/4}TaS_2$ [22]. However, a substantial DMI is allowed in $Fe_{1/3}TaS_2$ due to the combination of strong SOC of Fe and Ta atoms and the broken SIS [3-5], but absent in $Fe_{1/4}TaS_2$ due to the presence of SIS. Thus, the specific magnetic structure can be effectively described by a spin model,

$$H = -J\sum_{<i,j>} \mathbf{S_i} \cdot \mathbf{S_j} + \sum_{i,j} \mathbf{d_{ij}} \mathbf{S_i} \times \mathbf{S_j} + K\sum_i S_i^z \cdot S_i^z - B_z \sum_i S_i^z \qquad (1)$$

where the indices $i$ and $j$ sum over the Fe atoms. $J > 0$ is the FM exchange interaction, $\mathbf{d_{ij}}$ is the vector of DMI, $K < 0$ indicates PMA favoring an easy-axis, and the last term is the Zeeman energy due to the applied magnetic fields.

In terms of symmetry (point group $D_6$), the DMI are allowed $\mathbf{d_\perp}$ and $\mathbf{d_{//}}$, for the components perpendicular and parallel to the direction of the $\mathbf{c}$ axis [30]. The total DMI $\mathbf{d_{tot}}$ now can be written as $\mathbf{d_{tot}} = c_1 \mathbf{d_\perp} + c_2 \mathbf{d_{//}} = c_1(\mathbf{m_z} \frac{\partial \mathbf{m_x}}{\partial y} - \mathbf{m_x} \frac{\partial \mathbf{m_z}}{\partial y} -$

$\mathbf{m_z}\frac{\partial \mathbf{m_y}}{\partial x} + \mathbf{m_y}\frac{\partial \mathbf{m_z}}{\partial x}) + c_2(\mathbf{m_x}\frac{\partial \mathbf{m_y}}{\partial z} - \mathbf{m_y}\frac{\partial \mathbf{m_x}}{\partial z})$ , with arbitrary coefficients $c_{1,2}$ and reduced magnetization $\mathbf{m_{x,y,z}}$.

It is known that the large PMA would suppress the chiral conical order [31] or the chiral soliton phase [19], in which $\mathbf{d}_{//}$ twists the in-plane spin magnetic moments. In the work, we shall closely examine the DMI $\mathbf{d}_\perp$. We carry out the First-principles calculation of the DMI $\mathbf{d}_\perp$ ($\propto \Delta_E^{DMI}$) by evaluating the total energy differences between the clockwise and count-clockwise Bloch-type spin textures along lines of Fe atoms as shown in Fig. 4a [32]. The emergent THE at lower temperatures suggests that the DMI is sufficient to destabilize the FM state, forming the Bloch-type spin textures in bulk (Néel-type spin textures are probably ruled out, as discussed in Figure S12 in SI), such as skyrmions and chiral domain walls [33]. Due to the weak coupling among the TaS$_2$ layers, each layer of Fe atoms is an effective 2D magnetic system. Then, we first consider the 2D Bloch-type skyrmions. We further simulate the impact of gating by changing the electron number $N_e$ in the first-principles calculations. As shown in Fig. 4b, when the gate voltage increases (shifts the Fermi energy towards the negative values), the strength of DMI becomes larger. Besides the gate-induced change of carrier density, the protons intercalation may locally form proton concentration gradient which can also lead to the spatial inversion symmetry breaking and contribute DMI [34]. Specifically, the DMI at $E_F = -80\ meV$ is about twice of the value at $E_F = -0\ eV$. The topological Hall resistivity from 2D skyrmions is proportional to the strength of the emergent magnetic field $\mathbf{b_z}$, where $\mathbf{b_z} = n_s \phi_0 / 2\pi R^2$, with R being the diameter of skyrmion, $n_s$ being the density of skyrmions and $\emptyset_0 = hc/e$ being the flux quantization of emergent magnetic field of unit sphere for each skyrmion. Thus, with increasing of the strength of DMI, the density of 2D skyrmions increases greatly than linearly, leading to dramatic increase of THE. This increased density of skyrmion could facilitate the promising next generation low-energy and high-density storage spintronic devices based on skyrmion systems and the manipulation of dynamics of skyrmions through gate tunable DMI. Recently, the chiral domain wall with $Z_6$ vortex was suggested to account for the observation of THE in Mn$_3$Sn [35, 36]. Analogically, the chiral domain wall with $Z_6$ ($Z_2 \times Z_3$) vortex in Fe$_{1/3}$TaS$_2$ revealed by transmission electron microscopy [21] may also be a possible origin of large THE.

In conclusion, intercalating Fe atoms in a TMD 2H-TaS$_2$, Fe$_{1/3-\delta}$TaS$_2$ ($\delta \leq 0.05$) nanoplates exhibit large THE, demonstrating a strong transport evidence of sizable DMI and the emergence of Bloch-type spin textures (such as skyrmions and chiral domain walls). Gate-induced protons intercalation can largely modulate the amplitudes of topological Hall resistivities by approximately 420%, indicating the high tunability of DMI. Theoretical analysis and First-principles calculations suggest that the sizable DMI dominantly comes from the intercalated Fe atoms, playing a key role in magnetic orders and the formation of the chiral supercells with broken SIS. Our discovery demonstrates that dual-intercalation (intercalation of magnetic atoms and protons) is a promising way of tailoring DMI and manipulating chiral spin textures in 2H-TaS$_2$, greatly inspiring further investigations in a large family of TMDs.

# Methods

**Single Crystal Growth.** Single crystals of $Fe_xTaS_2$ were grown via chemical vapor transport method with iodine as the transport agent with suitable mole ratio and sealed in an evacuated quartz tube (Supplementary section 1).

**Device Fabrication and Transport Measurements.** Solid protonic electrolyte was prepared by the sol-gel processes by mixing tetraethyl orthosilicate (from Alfa Aesar), ethanol, deionized water, phosphoric acid (as a proton source, from Alfa Aesar, 85% wt%). The mixed solution was stirred and annealed before use (Supplementary section 2). Transport measurements were performed in a commercial Physical Property Measurement System (PPMS) with magnetic field up to $9\,T$. Protonic gating experiments were performed in commercial magnetic property measurement System (MPMS) with a maximal magnetic field of $7\,T$. To decrease the leaking current during the gating, voltage was swept at $250\,K$. Once the resistance was changed, the sample was quickly cooled down to low temperatures for magneto-transport measurements.

# Acknowledgements

The authors thank Yoichi Horibe, Lingyao Kong, Yusuke Masaki and Di Xiao for insightful discussions. This research was performed in part at the RMIT Micro Nano Research Facility (MNRF) in the Victorian Node of the Australian National Fabrication Facility (ANFF) and the RMIT Microscopy and Microanalysis Facility (RMMF). Work at RMIT university was supported by the Australian Research Council Centre of Excellence in Future Low-Energy Electronics Technologies (Project No. CE170100039). M. W., Y. Y., X. Z. and M. T. were supported by the NSF of China (Grants Nos. 11734003, and U1732274), the National Key R&D Program of China (Grant Nos. 2016YFA0300600, 2017YFA0303201, and 2017YFA0403502). X. Z. was supported by Youth Innovation Promotion Association of CAS (Grant No. 2017483). J. Z. was supported by the 100 Talents Program of Chinese Academy of Sciences (CAS) and also partially by the High Magnetic Field Laboratory of Anhui Province.

## Author information

### Contributions

L. W. and M. T. conceived the project. G. Z. fabricated the devices and performed the transport measurements, assisted by C. T., S. A., N. A., M. A., L. F. and M.-Y. W., J. Z., Y. Y. provided theoretical support. X. Z., J. W., M. W. grew the FGT crystals. G. Z., M.-Y. W., J. Z., M. T. and L. W. analysed the data and wrote the manuscript with assistance from all authors.

### Competing interests

The authors declare no competing financial or non-financial interests.

### Data availability

All data supporting the findings of this study are available from the corresponding author on request.

# References

1. Dzyaloshinsky, I. A thermodynamic theory of 'weak' ferromagnetism of antiferromagnetics. *J. Phys. Chem. Solids* **4**, 241-255 (1958).

2. Moriya, T. Anisotropic superexchange interaction and weak ferromagnetism. *Phys. Rev.* **120**, 91-98 (1960).

3. Nagaosa, N., Tokura, Y. Topological properties and dynamics of magnetic skyrmions. *Nat. Nanotechnol.* **8**, 899 (2013).

4. Kanazawa, N., Seki, S., and Tokura, Y. Noncentrosymmetric magnets hosting magnetic skyrmions. *Adv. Mater.* **29**, 1603227 (2017).

5. Fert, A., Reyren, N., Cros, V. Magnetic skyrmions: Advances in physics and potential applications. *Nat. Rev. Mater.* **2**, 17031 (2017).

6. Cheng, R., Okamoto, S. and Xiao, D. Spin Nernst effect of magnons in collinear antiferromagnets. *Phys. Rev. Lett.* **117**, 217202 (2016).

7. Zhang, X., Zhang, Y., Okamoto, S., and Xiao, D. Thermal Hall Effect Induced by Magnon-Phonon Interactions. *Phys. Rev. Lett.* **123**, 167202 (2019).

8. Park, S., and Yang, B. Topological magnetoelastic excitations in noncollinear antiferromagnets. *Phys. Rev. B* **99**, 174435 (2019).

9. Nagaosa, N. et al., Anomalous Hall effect. *Rev. Mod. Phys.* **82**, 1539-1592 (2010).

10. Ye, J. W. et al., Berry phase theory of the anomalous Hall effect: application to colossal magnetoresistance manganites. *Phys. Rev. Lett*. **83**, 3737-3740 (1999).

11. Bruno, P., Dugaev, V. K., and Taillefumier, M. Topological Hall Effect and Berry Phase in Magnetic Nanostructures. *Phys. Rev. Lett.* **93**, 096806 (2004).

12. Rößler, U. K., Bogdanov, A. N. & Pfleiderer, C. Spontaneous skyrmion ground states in magnetic metals. *Nature* **442**, 797-801 (2006).

13. Uchida, M., Onose, Y., Matsui, Y. & Tokura, Y. Real-space observation of helical spin order. *Science* **311**, 359-361 (2006).

14. Mühlbauer, S. et al. Skyrmion lattice in a chiral magnet. *Science* **323**, 915-919 (2009).

15. Kim, D. et al. Bulk Dzyaloshinskii–Moriya interaction in amorphous ferrimagnetic alloys. *Nat. Mater*. **18**, 685-690 (2019).

16. Friend, R. H., Beal, A. R., Yoffe, A. D. Electrical and magnetic properties of some first row transition metal intercalates of niobium disulphide. *Philos. Mag.* **35**, 1269−1287 (1977).

17. Parkin, S. S. P., Friend, R. H. 3d transition-metal intercalates of the niobium and tantalum dichalcogenides. I. Magnetic properties. *Philos. Mag.* **41**, 65–93 (1980).

18. Little, A. et al. Three-state nematicity in the triangular lattice antiferromagnet $Fe_{1/3}NbS_2$. *Nat. Mater.* (2020).

19. Togawa, Y. et al., Chiral magnetic soliton lattice on a chiral helimagnet. *Phys. Rev. Lett.* **108**, 107202 (2012).

20. Wang, L. et al. Controlling the Topological Sector of Magnetic Solitons in Exfoliated $Cr_{1/3}NbS_2$ Crystals. *Phys. Rev. Lett.* **118**, 257203 (2017).

21. Horib, Y. et al., Color Theorems, Chiral Domain Topology, and Magnetic Properties of $Fe_xTaS_2$. *J. Am. Chem. Soc.* **136**, 8368−8373 (2014).

22. Ko, K.-T. et al., RKKY Ferromagnetism with Ising-Like Spin States in Intercalated $Fe_{1/4}TaS_2$. *Phys. Rev. Lett.* **107**. 247201 (2011).

23. Yuan, H. et al., High-density carrier accumulation in ZnO field-effect transistors gated by electric double layers of ionic liquids. *Adv. Funct. Mater*. **19**, 1046-1053 (2009).

24. Lei, B. et al., Tuning phase transitions in FeSe thin flakes by field-effect transistor with solid ion conductor as the gate dielectric. *Phys. Rev. B* **95**, 020503(R) (2017).

25. Eibschütz, M. et al., Ferromagnetism in metallic $Fe_xTaS_2$ (x~0.28). *Appl. Phys.*

*Lett.* **27**, 464 (1975).

26. Dijkstra, J. et al., Band-structure calculations of $Fe_{1/3}TaS_2$ and $Mn_{1/3}TaS_2$, and transport and magnetic properties of $Fe_{0.28}TaS_2$. *J. Phys.: Condens. Matter* **1**, 6363-6379 (1989).

27. Hardy, W. J. et al., Very large magnetoresistance in $Fe_{0.28}TaS_2$ single crystals. *Phys. Rev B* **91**, 054426 (2015).

28. Zeng, C., Yao, Y., Niu, Q., Weitering, H. H. Linear Magnetization Dependence of the Intrinsic Anomalous Hall Effect. *Phys. Rev. Lett.* **96**, 037204 (2006).

29. Ohuchi, Y. et al., Electric-field control of anomalous and topological Hall effects in oxide bilayer thin films. *Nat. Commun.* **9**, 213 (2018).

30. Bogdanov, A., Hubert, A. Thermodynamically stable magnetic vortex states in magnetic crystals. *J. Magn. Magn. Mater.* **138**, 255 (1994).

31. Meynell, S. A. et al., Hall effect and transmission electron microscopy of epitaxial MnSi thin films. *Phys. Rev.* B **90**, 224419 (2014).

32. Yang, H., Thiaville, A., Rohart, S., Fert, A., and Chshiev, M. Anatomy of Dzyaloshinskii-Moriya Interaction at Co/Pt Interfaces. *Phys. Rev. Lett.* **115**, 267210 (2015).

33. Rohart, S., Thiaville, A. Skyrmion confinement in ultrathin film nanostructures in the presence of Dzyaloshinskii-Moriya interaction. *Phys. Rev. B* **96**, 037204 (2006).

34. Li, Z., Shen, S. et al., Reversible manipulation of the magnetic state in $SrRuO_3$ through electric-field controlled proton evolution. *Nat. Commun.* **11**, 184 (2020).

35. Liu, J., Balents, L. Anomalous hall effect and topological defects in antiferromagnetic weyl semimetals: $Mn_3Sn/Ge$. *Phys. Rev. Lett.* **119**, 087202 (2017).

36. Li, X. et al., Chiral domain walls of $Mn_3Sn$ and their memory. *Nat. Commun.* **10**, 3021 (2019).

# Figure caption

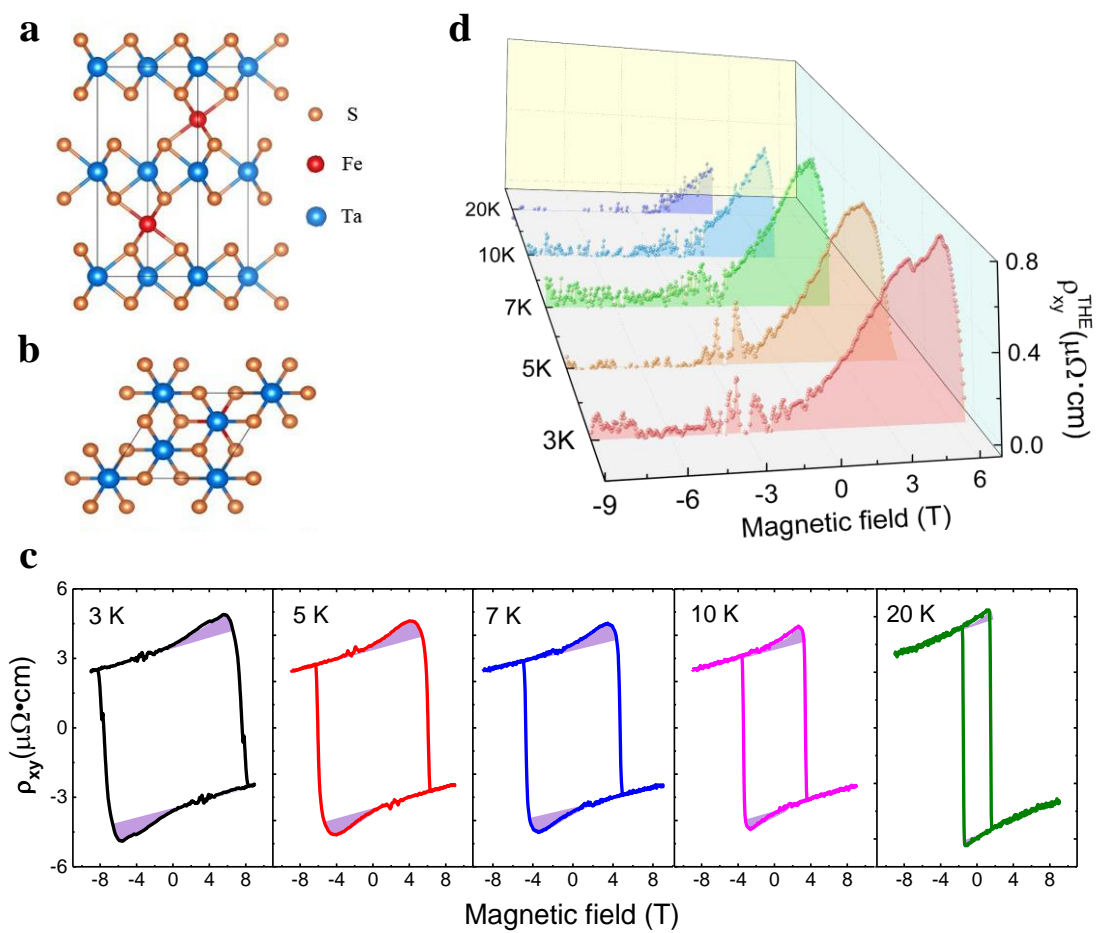

**Fig. 1** Topological Hall effects observed in Fe-intercalated 2H-TaS$_2$. **a, b** Crystal structures of Fe$_{1/3}$TaS$_2$ of front view (a) and top view (b). **c** Temperature-dependent Hall resistivity $\rho_{xy}$ in Fe$_{0.28}$TaS$_2$ nanoflakes. Topological Hall resistivity components $\rho_{xy}^T$ are shadowed by the light purple color. **d** Temperature-dependent Topological Hall resistivity components $\rho_{xy}^T$.

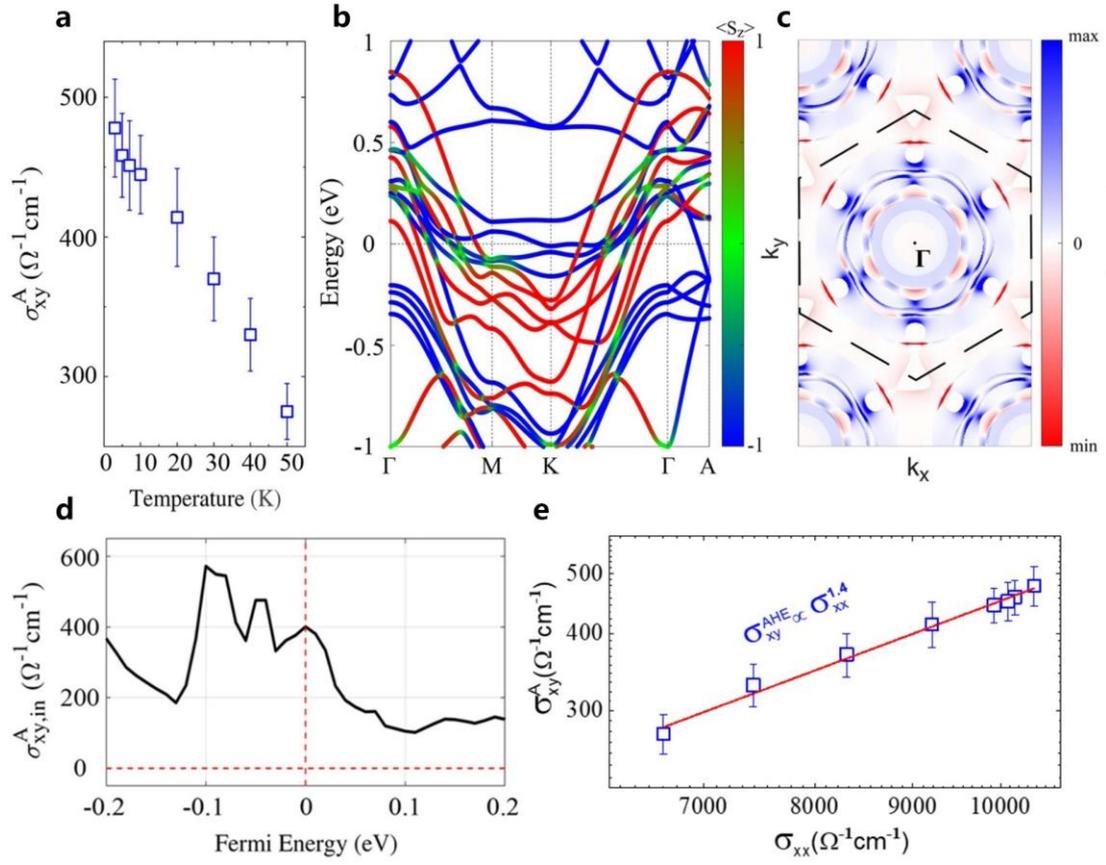

**Fig. 2** Anomalous Hall conductivity and nontrivial band structure. **a** Temperature-dependent anomalous Hall conductivity in S1. **b, c** Band-structure of Fe$_{1/3}$TaS$_2$. The colours mark the spin expectation $\langle S_z \rangle$ of the band. **d** Berry curvature distribution in $k_z = 0$ plane. **d** Fermi Energy $E_F$ dependent intrinsic AHC $\sigma^A_{xy,in}$. **e** Scaling relationship between anomalous Hall conductivities $\sigma^{AHE}_{xy}$ and longitudinal conductivities $\sigma_{xx}$.

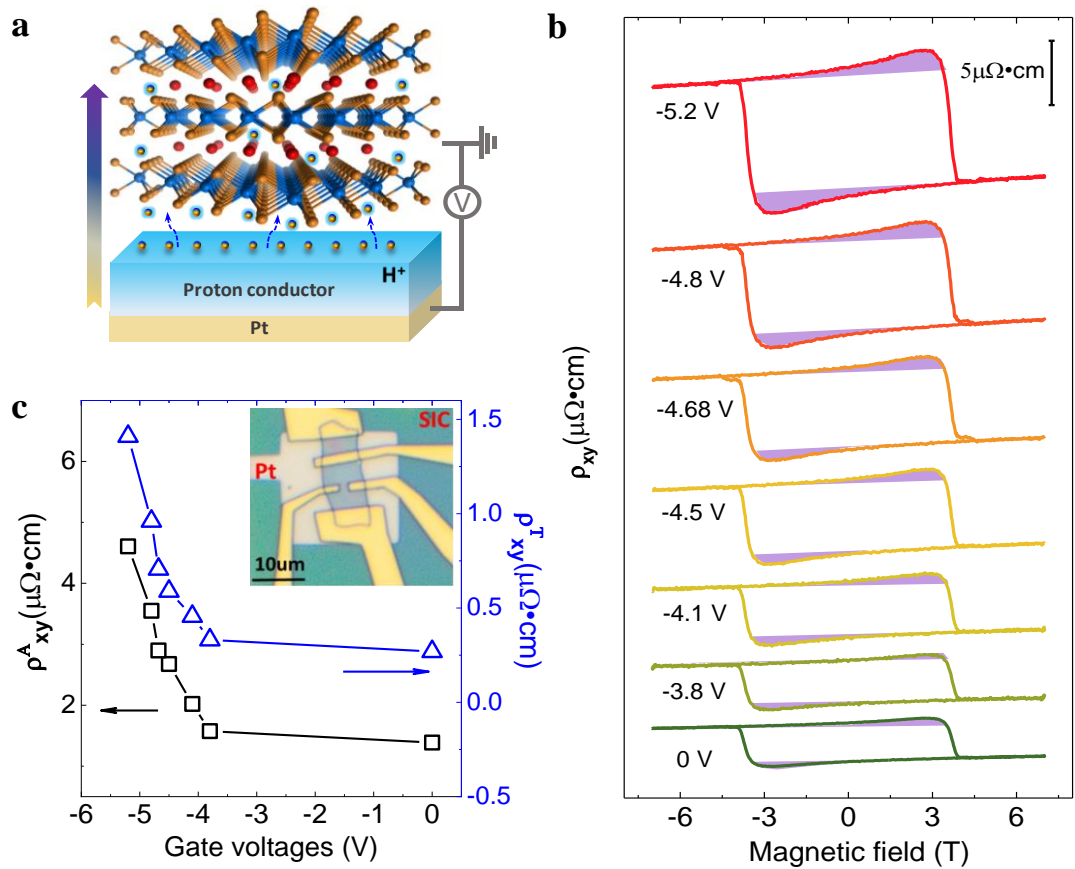

**Fig. 3** Gate-tuned anomalous Hall and topological Hall effect in $Fe_{0.28}TaS_2$ nanoflake. **a** Schematic of gate-induced proton intercalations. **b** Hall resistivity under different gate voltages. Topological Hall resistivity $\rho_{xy}^{T}$ are shadowed by the light purple colour. **c** Gate-dependent anomalous and topological Hall resistivity. Inset: a Hall-bar device on solid ion (proton) conductor (SIC).

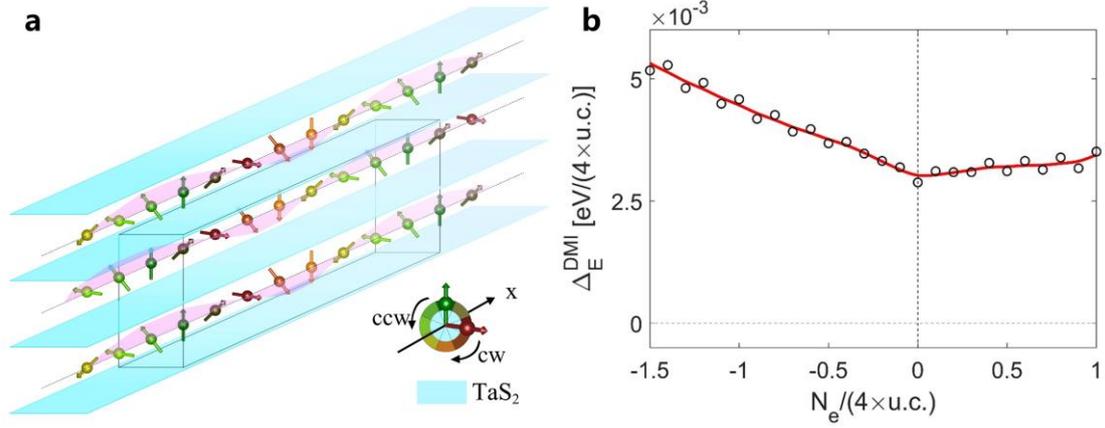

**Fig. 4** DMI simulation in First-principles calculations. **a** Spin configurations used to calculate DMI strength. Spins are represented by arrows. **b** The electron number $N_e$ dependent DMI strength $\mathbf{d}_\perp$ ($\propto \Delta_E^{DMI}$) for the supercell with four-spin cycle along one selected direction in **a**. $N_e = 0$ represents the case of $E_F = 0$, and $N_e = -1.5(+1)$ represents the Fermi energy around $-80\ meV\,(+50\ meV)$. The black circles are the results of First-principles calculation and the red line is fitted from the black circles for guiding eyes.